# Experimental Investigation of Discrete Multitone Transmission in the Presence of Optical Noise and Chromatic Dispersion

**Annika Dochhan[1], Laia Nadal[2], Helmut Griesser[3], Michael Eiselt[1], Michela Svaluto Moreolo[2], and Jörg-Peter Elbers[3]**

1 ADVA Optical Networking SE, Maerzenquelle 1-3, 98617 Meiningen, Germany
2 CTTC, Av. Carl Friedrich Gauss 7, Castelldefels, Spain
3 ADVA Optical Networking SE, Fraunhoferstr. 9a, 82152 Martinsried, Germany
ADochhan@advaoptical.com

**Abstract:** Enabled by channel adaptive bit and power loading, we experimentally demonstrate discrete multitone transmission at 56Gb/s with simple intensity modulation and direct detection and achieve 50 km reach in the 1.55μm window.

## 1. Introduction

Discrete Multitone Transmission (DMT) with intensity modulation (IM) and direct detection (DD) has recently emerged as promising candidate for 100 GbE interconnects [3]. As variant of orthogonal frequency division multiplexing (OFDM) with real valued time signal, it also offers a potentially low-cost approach for N x 100Gb/s dense wavelength division multiplexing (DWDM) inter-data center interconnects over distances beyond 40 km. While conventional DMT easily can compensate high values of chromatic dispersion (CD), the DD of a standard double sideband (DSB) signal transforms the optical channel into a highly frequency selective fading channel for large amounts of CD, thereby limiting the usable bandwidth when operated in the 1.55μm wavelength range (as required for DWDM transmission). The standard approach that allows even long-haul transmission with DMT systems [1] uses a single sideband (SSB) signal to prevent the power fading problem. A frequency gap equal to the bandwidth of the electrical signal is necessary between the carrier and the first modulated subcarrier to avoid signal-signal beat distortions in that frequency region. Drawbacks of this approach are the increased complexity of the setup due to the need for a single sideband filter or optical I-Q-modulator as well as higher bandwidth requirements of the components. Milion [2] did an early experimental demonstration of the transmission of a DSB DMT system with bit and power loading, achieving 19 Gb/s over a 25 km PON link. A later experimental study by Yan [3] showed 100 Gb/s over a distance of 10 km and 60 Gb/s over 40 km. For the application of the technique to optically amplified links that allow longer distances, Paul [4] suggested a simplified 'on-off' loading for 42.8 Gb/s over a 80 km span with optical pre-amplification, showing by means of simulations that loading can be an adequate alternative to the 'gap approach'. Also, in a detailed simulative study [5], various DMT variants were compared with optimized bit (BL) and power loading (PL).

In this paper, we experimentally demonstrate the optical transmission of 56 Gb/s DSB DMT (including 7% FEC overhead) without any explicit frequency gap over a dispersive and optically amplified link of up to 80 km of SSMF at 1.55μm. To the best of our knowledge this is the highest data rate for such a system achieved over this distance. The required optical bandwidth is around 45 GHz, thus offering the prospect of operation over DWDM links with 50 GHz channel grid. Combining 2 or 8 optical carriers to a superchannel, multiple 100 Gb/s or 400Gb/s signals can be transmitted.

## 2. Experimental setup

The experimental setup is shown in Figure 1. The electrical DMT signal was generated offline using Python routines and stored inside the memory of a digital to analog converter (DAC) with 64 GS/s. It was amplified by a linear driver and modulated onto the optical carrier at 192.5 THz by a Mach-Zehnder-Modulator (MZM). The MZM was biased at the power quadrature point, leading to a strong carrier, but reduced subcarrier-subcarrier intermodulation products. The DAC exhibits a bandwidth of ~ 13 GHz and 64 GS/s while the driver and the MZM are suited for 30 GHz and 40 GHz bandwidth signals, respectively. Subsequently the optical signal was transmitted over a single span of standard single mode fiber (SSMF) with 50.5 or 82.1 km length and a loss of 0.2 dB/km. The launch power into the SSMF was set to 5 dBm. At the receiver, after optional additive noise loading to control the OSNR, the signal was optically pre-amplified by an Erbium-doped fiber amplifier (EDFA). After noise filtering with a 100 GHz demultiplexer, the signal was directly detected with a 50 GHz bandwidth PIN photo diode and captured by an 80 GS/s real time oscilloscope with 29.4 GHz bandwidth. Offline processing was again performed using Python. Due to component availability, the devices used in this setup have a bandwidth well beyond 20 GHz. However, since the bandwidth requirement is mainly determined by the DAC and the DD optical channel, optoelectronic devices with a bandwidth below 20 GHz could have been used, resulting in a more cost-effective design.

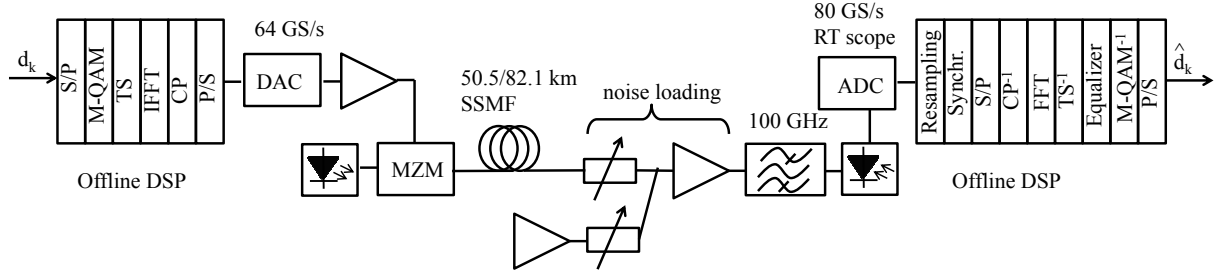

Fig. 1: Experimental setup and signal processing steps for the DMT transmission system under investigation.

Fig. 1 also shows the main DSP building blocks, which are needed in order to obtain the DMT signal to be transmitted from a random data sequence $d_k$. The BL and PL distributions were determined by sending a probe signal with a uniform 16-QAM bit loading for all subcarriers, estimating the signal-to-noise ratio (SNR) at the receiver side and finally applying the Levin-Campello loading algorithm [6]. For symbol synchronization and channel estimation five training symbols (TS) are inserted. Limited by the DAC memory, the total length of a DMT frame is 124 symbols (including TS). We used a 2048-point IFFT, yielding 1023 usable carriers (no data is transmitted on the central subcarrier) to obtain a real-valued baseband signal. Out of these, at maximum 852 carriers are actually modulated to introduce some oversampling to support aliasing filtering. A cyclic prefix of 1/64 was used, leading to data rates of 56 Gb/s to 112 Gb/s (including TS). The DMT time signal was clipped with a clipping ratio of 12 dB to reduce the peak-to-average power ratio. At the receiver side the symbols can be properly de-mapped for Monte Carlo type error counting after applying resampling, synchronization (a variant of the Schmidl-Cox algorithm), FFT, removal of all overhead, and one tap equalization with decision directed channel estimation.

## 3. Results

In a first measurement, the system was operated without optical fiber to assess the back-to-back OSNR performance of the modulation format. Fig. 2 (left) shows the bit error ratio (BER) versus the optical signal-to-noise ratio (OSNR). For an error correcting threshold of 4e-3 (assuming a hard decision forward error correcting (HD-FEC) scheme with 7% overhead), error free performance requires 31 dB OSNR for 56 Gb/s and 40 dB for 96 Gb/s. A rate of 112 Gb/s shows an error floor above the HD-FEC threshold. Using a soft decision FEC (SD-FEC, threshold 1.9e-2) this gross data rate would be still feasible at the cost of a higher overhead and thus reduced net bit rate. The right side of Fig. 2 shows the BL distribution with the estimated SNR as an inset.

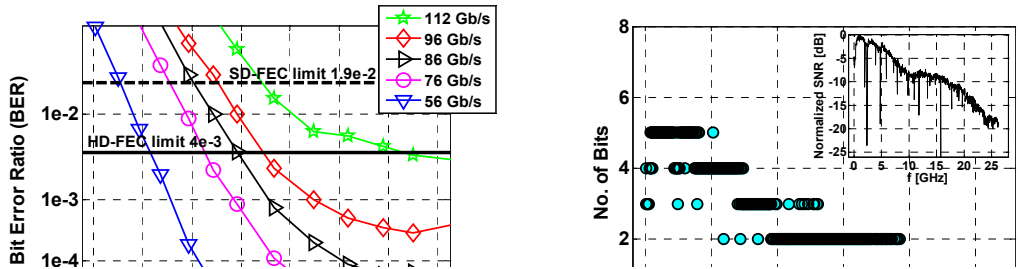

Fig. 2: OSNR sensitivity of the DMT system for back-to-back (left), the result of the bitloading for 56 Gb/s (right), and the estimated SNR (inset) that is the input for the bit and power loading algorithm.

After the transmission over 50.5 km of SSMF (see Fig. 3, left) a data rate of 56 Gb/s requires an increased OSNR of 37 dB and 76 Gb/s is no longer feasible with HD-FEC. An SD-FEC, however, would allow error-free operation at an OSNR of 40 dB. The chromatic fiber dispersion leads to significant power fading in the frequency region of interest, as is visible from the SNR estimation in the inset of Fig. 3 (right), and the BL algorithm has to assign more bits to lower subcarriers as can be seen in Fig. 3 (right). The resulting use of constellations up to 128-QAM is the main reason for the loss in OSNR sensitivity compared to back-to-back.

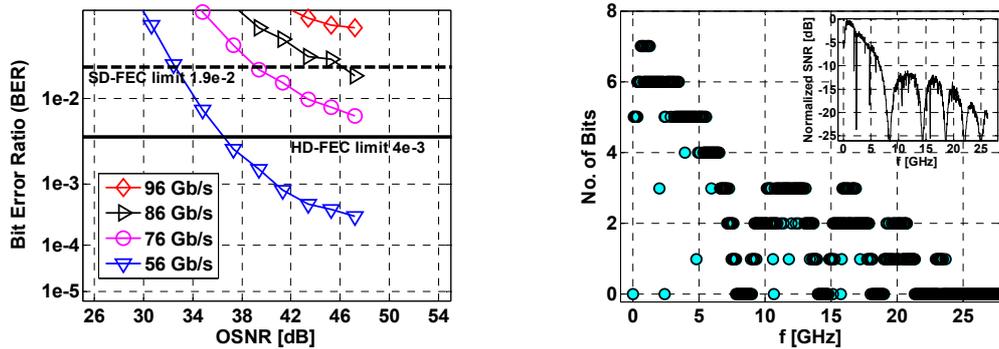

Fig. 3: OSNR sensitivity of the DMT system after transmission over 50.5 km of SSMF (left), the result of the bitloading for 56 Gb/s (right), and the estimated SNR (inset).

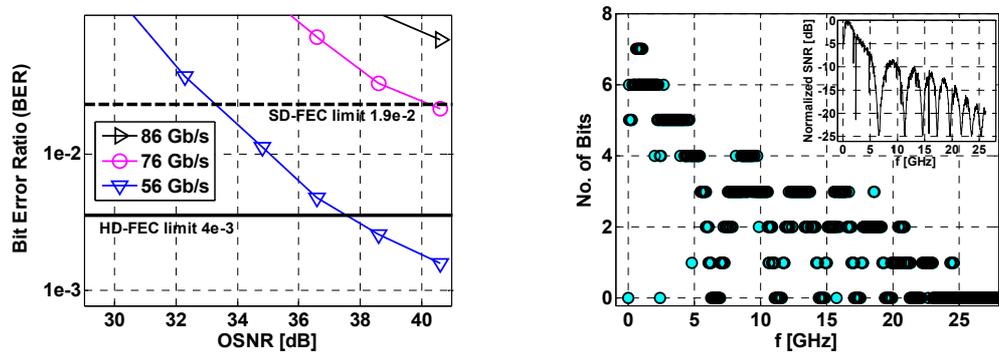

Fig. 4: OSNR sensitivity of the DMT system after transmission over 82.1 km of SSMF (left), the result of the bitloading for 56 Gb/s (right), and the estimated SNR (inset).

In a third experiment the fiber span was increased to 80.1 km. Fig. 4 (left) shows the results for data rates of 56, 76, and 86 Gb/s. For a data rate of 56 Gb/s, the required OSNR of 38 dB is only slightly increased compared to the 50 km transmission, whereas for 76 Gb/s only an SD-FEC would allow error-free transmission. The BL shows a similar pattern as for the 50 km span with even more frequent fading, slightly increasing the number of high order constellations.

### 4. Conclusions
Using discrete multitone modulation, we demonstrated 56 Gb/s optical transmission in the 1.55μm window over up to 80 km of SSMF without optical dispersion compensation with an error rate below the threshold of a 7% overhead hard decision FEC. This performance is achieved by employing a per-subcarrier bit and power loading that is adapted to the dispersive channel with optical noise. By avoiding the frequency gap, we believe that the concept is suitable for DWDM systems with a 50 GHz grid. Combining 2 or 8 DMT signals to optical superchannels, it allows to transmit multiple 100 or 400Gb/s signals over high capacity DWDM links as used for high capacity inter-datacenter interconnects.

### 5. Acknowledgements
The results were obtained in the framework of the SASER-ADVAntage-NET project, partly funded by the German ministry of education and research (BMBF) under contract 16BP12400, the FP7 EU-Japan project STRAUSS (GA 608528), and the project FARO (TEC2012-38119), the FPI research scholarship grants BES-2010-031072 and EEBB-I-13-06102 funded by the Spanish MINECO.

### 6. References

[1] B.J.C. Schmidt, A.J. Lowery, J. Armstrong, JLT, Vol. 26, No. 1, (2008), pp. 196-203
[2] C. Milion *et al.,* ECOC'09, paper 7.5.4.
[3] W. Yan *et al.*, OFC'13, OM3H.1
[4] H. Paul and K.-D. Kammeyer, ECOC'09, paper P3.11
[5] D.J.F. Barros and J.M. Kahn, JLT, Vol. 28, No. 12 (2010), pp. 1811-20
[6] J. M. Cioffi, "Data Transmission Theory," course text for EE379C (http://www.stanford.edu/group/cioffi/).